\documentclass[11pt,twoside]{article}

\usepackage{asp2010}
\usepackage{graphicx}

\resetcounters 

\begin{document}

\resetcounters

\title{Closed loop simulations of the thermal experiments in LISA Pathfinder}


\author{Ferran Gibert, Miquel Nofrarias, Nikolaos Karnesis, Marc D\'iaz-Aguil\'o, Ignacio Mateos, Alberto Lobo, Llu\'is Gesa, V\'ictor Mart\'in, and Ivan Lloro 
\affil{Institut de Ci\`encies de l'Espai ICE-CSIC/IEEC, Barcelona, Spain}}
 
\begin{abstract} 
The thermal experiments to be carried out onboard LISA Pathfinder (LPF) will provide essential information of the dependences of the instrument with respect to temperature variations. These thermal experiments must be modelled and simulated both to be validated for mission operations purposes and also to develop a data analysis tool able to characterise the temperature noise contribution to the instrument performance. Here we will present the models developed and the simulated signals for some of the experiments together with the corresponding interferometer readouts, the latter being computed by combining the thermal models with the global LTP (LISA Technology Package) simulator of the LTP Data Analysis team.
\end{abstract}

\section{Introduction} 

The performance of the LTP interferometers (IFO) onboard LISA Pathfinder~\citep{Antonucci:2010wm} must reach an extremely low frequency differential acceleration sensitivity:
\begin{eqnarray}
  	\label{reqeq}
    S^{1/2}_{\Delta a,\, {\rm LPF}}(\omega)\leq 3\times10^{-14}\left[1+\left(\frac{\omega/2\pi}{3\, {\rm mHz}}\right)^{2}\right]\, {\rm m}\, {\rm s}^{-2}\, {\rm Hz}^{-1/2} 
  \end{eqnarray}
  in the bandwidth between 1\,mHz and 30\,mHz. \\

At such low frequencies, the temperature noise shows up to be one of the potential perturbations threatening the final sensitivity of the instrument. Such temperature noise cause effects on the system through mainly three different thermal-sensitive elements~\citep{thermalDiag}: 

\begin{enumerate}
\item{{\it Test Masses of the Inertial Sensors:} Asymmetric temperature distributions around the Test Masses (TM) inside the Electrode Housing (EH) create forces and torques on them through different thermal effects, as the radiometer effect, the radiation pressure effect and the asymmetric outgassing~\citep{trentopaper}.}
\item{{\it Optical Windows:} Temperature fluctuations on the Optical Window's glass cause phase shifts to the laser beam~\citep{owpaper}.}
\item{{\it Optical Bench:} Thermo-elastic distortions of the Optical Bench (OB) affect the IFO signal~\citep{toqmdoc}.}
\end{enumerate}

\begin{figure}[!h]
	\begin{center}
		\includegraphics[width=26pc]{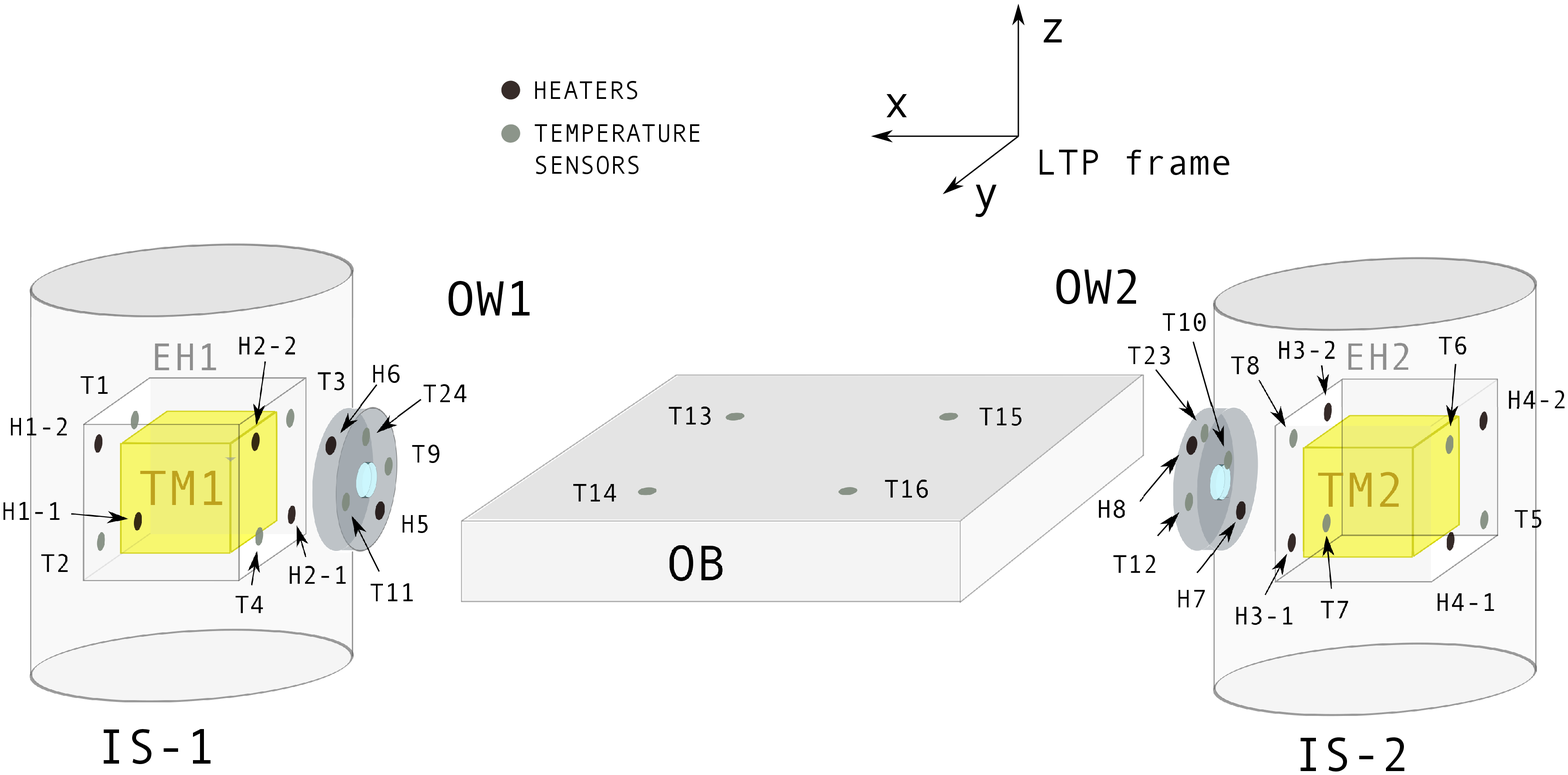}
		\caption{\label{dds_items} Thermal diagnostics items for the Inertial Sensor's, Optical Windows and Optical Bench.}
	\end{center}
\end{figure}

Therefore, the thermal diagnostic system was conceived to characterise the influence of temperature perturbations at such elements on the IFO readouts by means of different experiments, by injecting sequences of heat pulses at specific points of the LTP. Figure~\ref{dds_items} presents the layout of thermal diagnostic items concerning the Inertial Sensor's Electrode Housings, the Optical Windows (OW) and the Optical Bench (OB), and consisting of a set of heaters and high precision temperature sensors.

Models for each experiment are being developed in order to optimise the input signals but also as a tool for future data analysis during the mission~\citep{procAmaldi9Ferran}. In this contribution, we present for the first time the expected interferometer signals concerning the EH and the OW diagnostics experiments. Expected signals from the struts heating experiments to study the thermo-elastic distortions of the OB are not presented since their models, which will be based on data from the Thermo-Optical Qualification Model (TOQM) campaign~\citep{toqmdoc}, are still under development.

\section{Model description}

The modelling of each thermal experiment consists mainly of two parts: the first one, common to all of them, contains the thermal information of how heat propagates through the system, and it delivers time series of temperatures at specific points after the activation of a set of heaters. This block is based on the data extracted from a complete ESATAN 
thermal model of the whole spacecraft.

The second part is specific for each kind of thermal experiment as either it models the different EH thermal effects or it implements digital filters for the OW case. Finally, time series of both forces and torques for the EH thermal experiments and IFO phase shift time series for the OW experiments are obtained. More details of their modelling are described in~\citet{procAmaldi9Ferran}.

\begin{figure}[!h]
	\begin{center}
		\includegraphics[width=25pc]{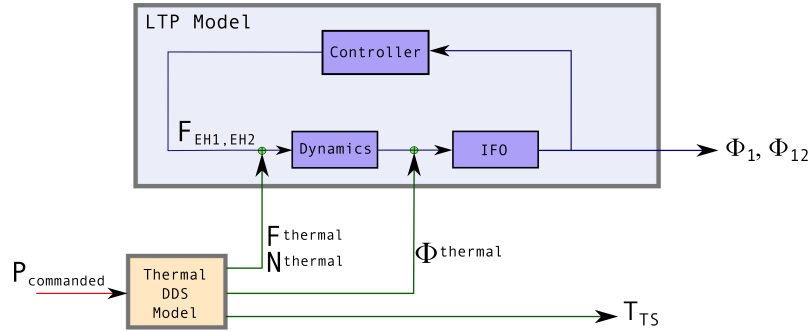}
		\caption{\label{loopscheme} Simplified scheme of the LTP control loop. Notice the injection points for the inputs forces and torques ($F^{thermal}$, $N^{thermal}$ respectively) before the Dynamics block and the phase shift ($\Phi^{thermal}$) after it.}
	\end{center}
\end{figure}

All these models have been implemented as state-space models under the frame of the LTPDA Toolbox~\citep{ltpdaref}, which is a MATLAB 
Toolbox being developed by the LTP Data Analysis team. Such a Toolbox is intended to provide a common framework for all the data analysis operations during the mission, and it is also being used to develop a detailed simulator of the LTP aimed to provide expected signals and other relevant information of the LTP behaviour prior to the mission. The scheme of the simulator loop with the injection points for the different time series is shown in Figure~\ref{loopscheme}.

\section{Results}
The simulations presented here have been done considering typical input signals. For each experiment, the input consists of:
\begin{itemize}
\item {\it EH experiments:} Two pulses of 10\,mW and a length of 1000\,s applied alternately to H1 and H2, producing a 0.5\,mHz temperature gradient signal on the X axis of TM1 of two whole cycles.
\item {\it OW experiments:} Five pulses of 0.5\,W simultaneously to H5 and H6 (1\,W in total). The length of the pulses is 50\,s and they are applied after every 1000\,s.
\end{itemize}

\begin{figure}[!h]
	\begin{center}
		\includegraphics[width=15.7pc]{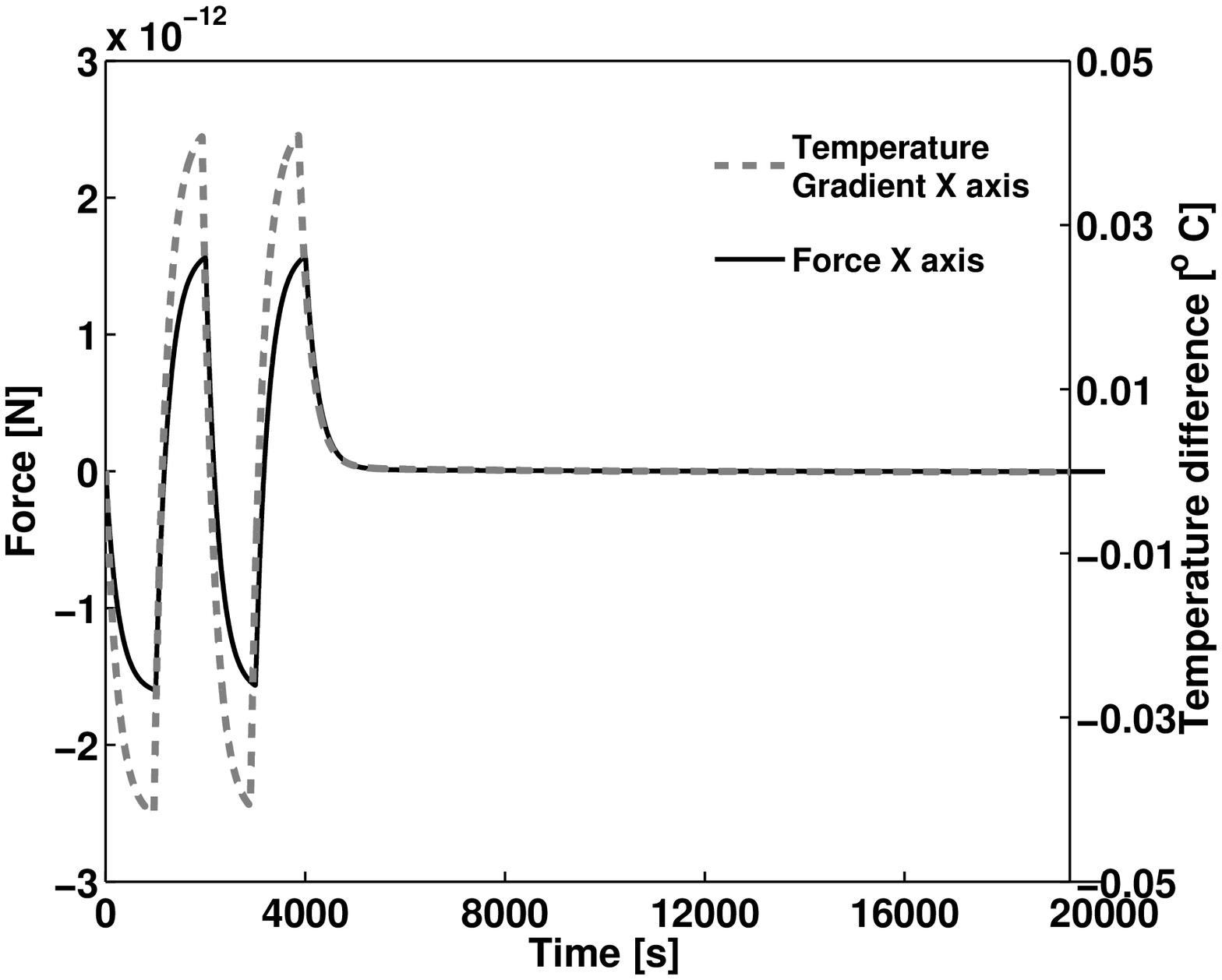}
		\includegraphics[width=15.7pc]{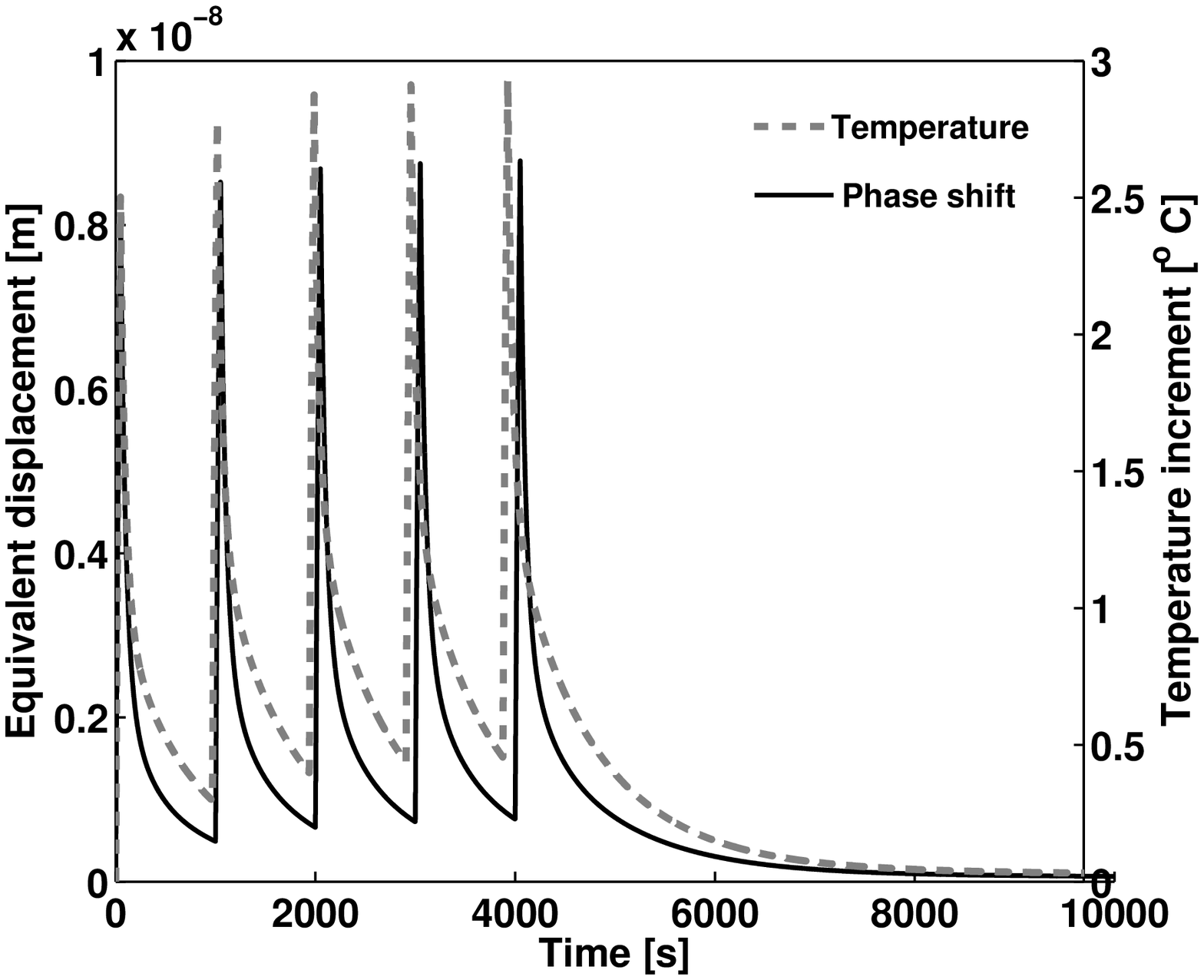}
		\caption{\label{Temps} {\it Left:} Test Mass 1 experienced force on the X axis during EH heating simulation, where linear dependence can be observed. {\it Right:} Phase shift and temperature variation during OW experiments.}
	\end{center}
\end{figure}

The consequences of such heat inputs are direct effects on the temperature at nearby spots. For the case of the EH experiments, the temperature gradient experienced and the corresponding force created is presented in Figure~\ref{Temps} ({\it left}). The quasi-linear dependence between forces and temperature gradient explained in~\citep{procAmaldi9Ferran} can be observed. 

With respect to the OW heating experiment, Figure~\ref{Temps} ({\it right}) shows the IFO phase shift in front of the temperature variation. In this case the dependence, as found in~\citep{owpaper}, has a low pass filter behaviour.

\begin{figure}[!h]
	\begin{center}
		\includegraphics[width=15.7pc]{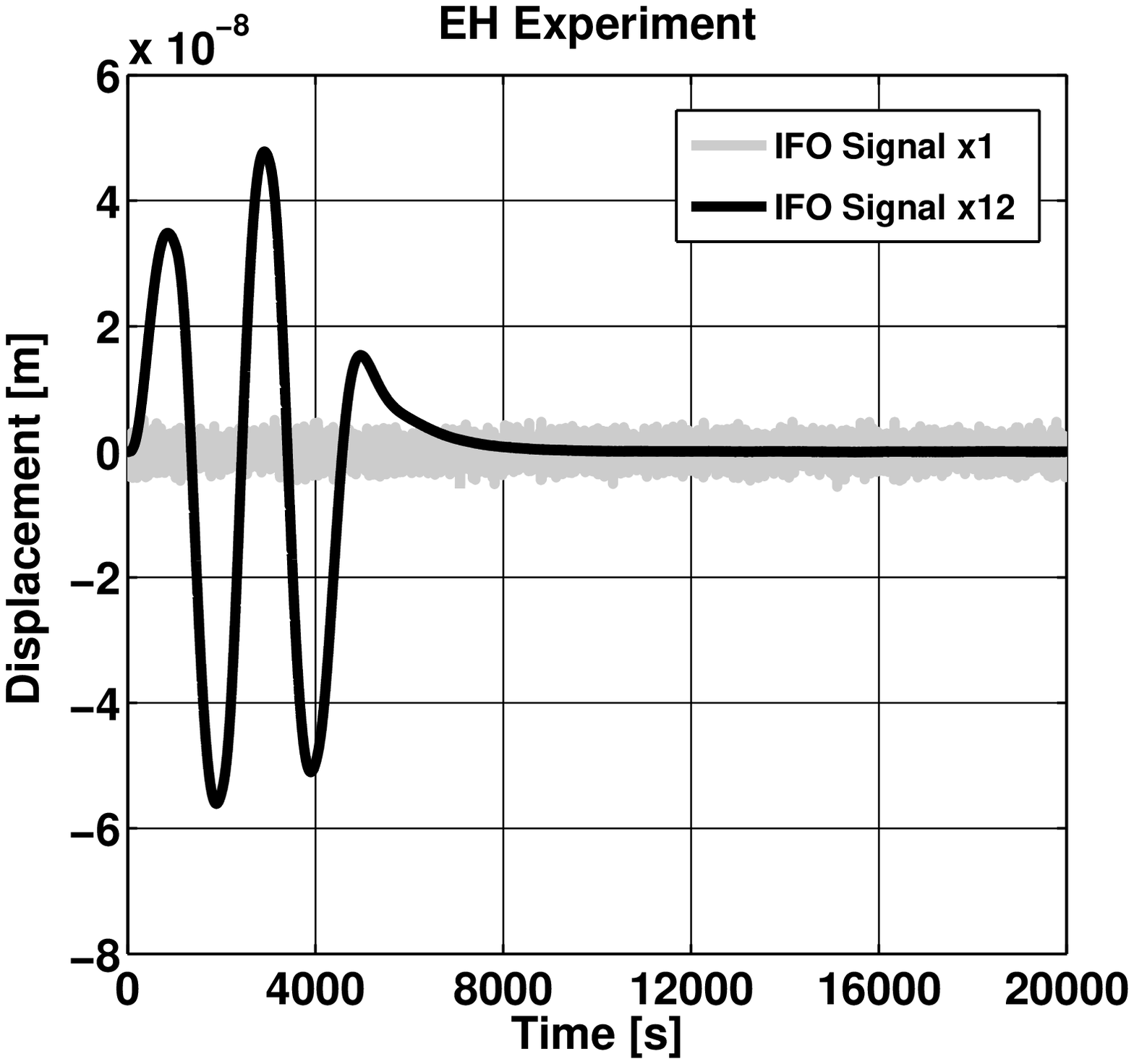}
		\includegraphics[width=15.7pc]{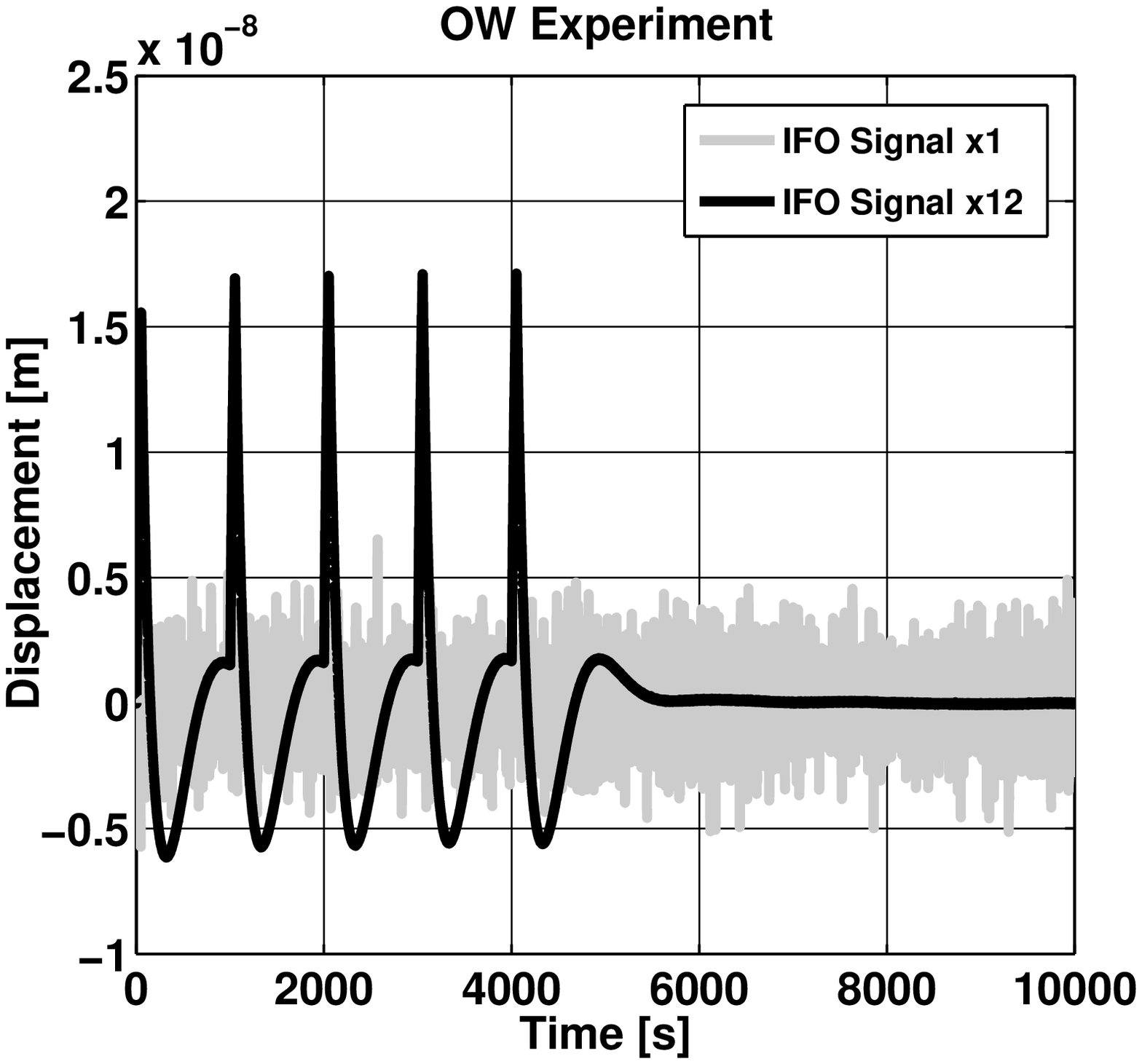}\\
		\caption{\label{ifos} { {\it Left:} IFO readout in EH experiment. {\it Right:} IFO readout in OW experiment.}}
	\end{center}
\end{figure}

\begin{figure}[!h]
	\begin{center}
		\includegraphics[width=15.7pc]{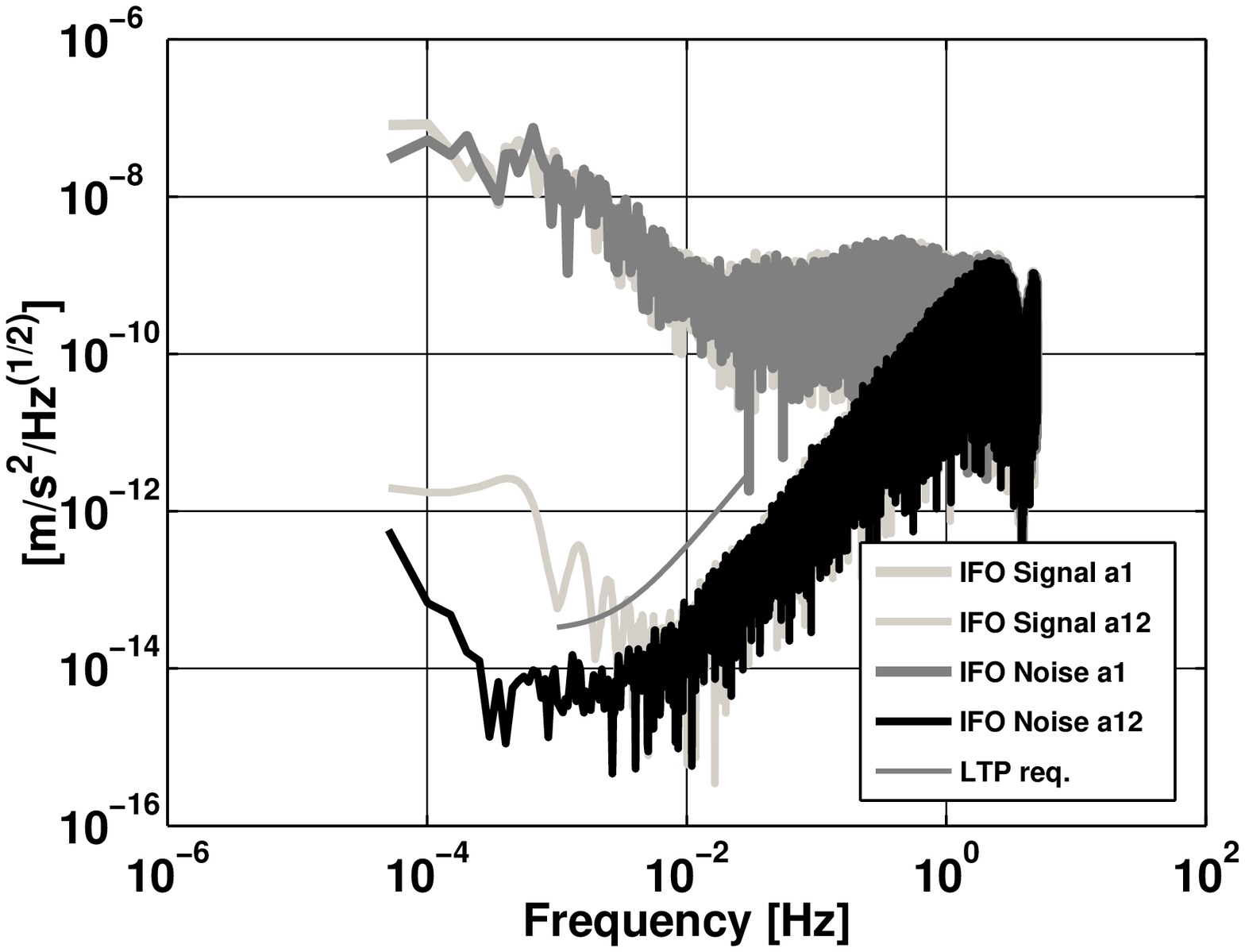}
		\includegraphics[width=15.7pc]{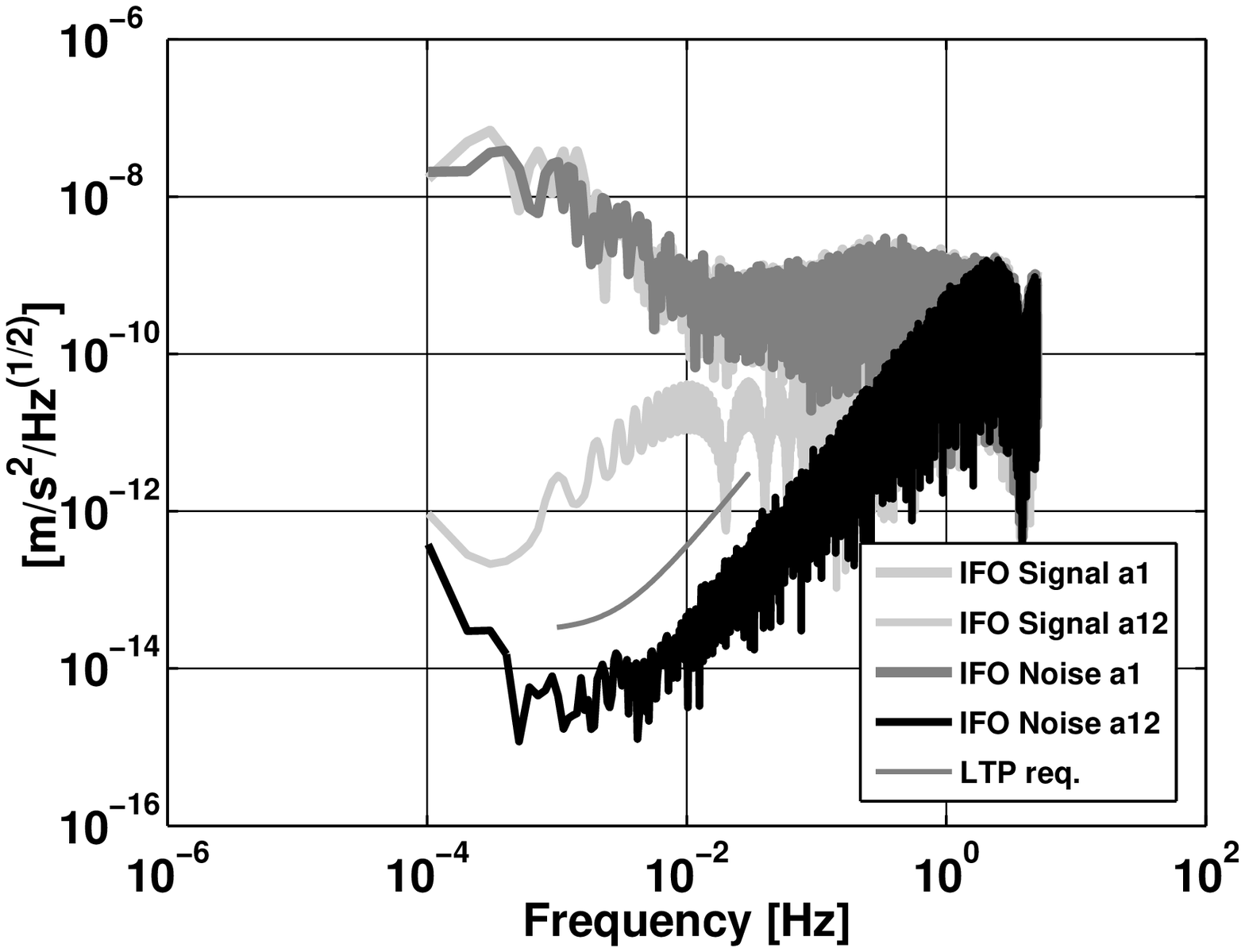}
		\caption{\label{ifos2} {{\it Left}: Accelerations ASD derived from IFO readouts in EH experiments. {\it Right:} Idem for OW experiments. Noise signals are obtained by injecting white noise temperature of $10^{-4}\,{\rm K/Hz^{1/2}}$ between 1\,mHz and 30\,mHz to the model inputs, considering common spectra profiles for the nodes on same EH walls.}}
	\end{center}
\end{figure}

Final IFO readouts for each experiment are presented in Figures~\ref{ifos}~and~~\ref{ifos2}. The time series observed represent real distance, while the acceleration amplitude spectral density is derived from the different IFO readouts.


As regards the IFO time series plots, the impact on the differential channels ({\it x12}) for both EH and OW experiments is clearly observed, as the pulses are evident. However, in the {\it x1} channel (distance between TM1 and the spacecraft) the signal injected is lost, as the control loop that makes the spacecraft follow TM1 operates at high frequencies and consequently the signal effect is lower than the total noise.

About the amplitude spectral density plots, in both EH and OW cases the signals applied produce noticeable effects on IFO differential channels, exceeding by more than one order of magnitude in general the LTP requirements and the expected noise level. Such results look reasonable, since a {\it SNR} of 50-100 is expected for thermal signals. On the other hand, this excess is total in the OW case and partial in the EH case, the latter producing effects only on the lowest part of the bandwidth. Regarding the {\it a1} channel, no important differences are appreciated between the signal and the noise injection simulations for both EH and OW cases, as expected from the time series plots.

\section{Summary}

Models for the simulation of both EH and OW thermal experiments for LISA Pathfinder have already been developed under the LTPDA frame and injected to a global LTP simulator. Such modelling has required the assembly of data from ESATAN thermal models and their conversion to state-space models, allowing fast simulations of typical LPF Thermal diagnostics experiments.

With current models, the closed-loop performance of the LTP during these experiments (EH and OW cases) can now be simulated, obtaining quite satisfactory results. However, some parameters of the EH thermal model still need to be tuned with information of on-ground experiments like the Torsion Pendulum of the University of Trento~\citep{trentopaper}. 

Future work will include the addition of models for the thermal diagnostics experiments involving the heating of the different struts. To do so, data from the recent TOQM campaign is going to be used to extract transfer functions.

\acknowledgements We acknowledge support from Project AYA2010-15709 of Plan Nacional del Espacio of the Spanish Ministry of Science and Innovation (MICINN).

\bibliography{fgibert_MyBibli}
\bibliographystyle{asp2010}

\end{document}